\documentclass[doublecol]{epl2} 

\usepackage{graphicx}
\usepackage{verbatim}
\usepackage{times}
\usepackage{amssymb}

\begin{document}

\title{Sound propagation and force chains in granular materials}
\author{Eli T. Owens \and Karen E. Daniels\thanks{ E-mail: kdaniel@ncsu.edu}}
\institute{Department of Physics, NC State University, Raleigh, NC, USA 27695}
\date{22 July 2010}


\abstract{Granular materials are inherently heterogeneous, leading to challenges in formulating accurate models of sound propagation. In order to quantify acoustic responses in space and time, we perform experiments in a photoelastic granular material in which the internal stress pattern (in the form of force chains) is visible. We utilize two complementary methods, high-speed imaging and piezoelectric transduction, to provide particle-scale measurements of both the amplitude and speed of an acoustic wave in the near-field regime. We observe that the wave amplitude is on average largest within particles experiencing the largest forces, particularly in those chains radiating away from the source, with the force-dependence of this amplitude in qualitative agreement with a simple Hertzian-like model of particle contact area. In addition, we are able to directly observe rare transient force chains formed by the opening and closing of contacts during propagation. The speed of the leading edge of the pulse is in quantitative agreement with predictions for one-dimensional chains, while the slower speed of the peak response suggests that it contains waves which have travelled over multiple paths even within just this near-field region. These effects highlight the importance of particle-scale behaviors in determining the acoustical properties of granular materials.}

\maketitle

\section{Introduction}

Sound propagation in granular materials differs from propagation in ordinary elastic materials in that there is a poor separation of length scales, particularly manifest in the branching networks of force chains which transmit stresses between particles. Continuum models \cite{Digby-1981-EEM, Goddard-1990-NEP, Velicky-2002-PDS}, including effective medium theory (EMT), have failed to quantitatively describe important features such as the dependence of the sound speed on pressure \cite{Jia-1999-UPE, Makse-1999-WEM, Makse-2004-GPN}. Particle-scale changes in the coordination number and force chains are likely responsible for important deviations from these models, and have been the focus of numerical simulations on model amorphous systems near jamming, where soft modes are important \cite{Zeravcic-2008-LBV, Xu-2009-ETJ, Wyart-2010-SPT}.

\begin{figure*}
\centering
\includegraphics[width=1.8\columnwidth]{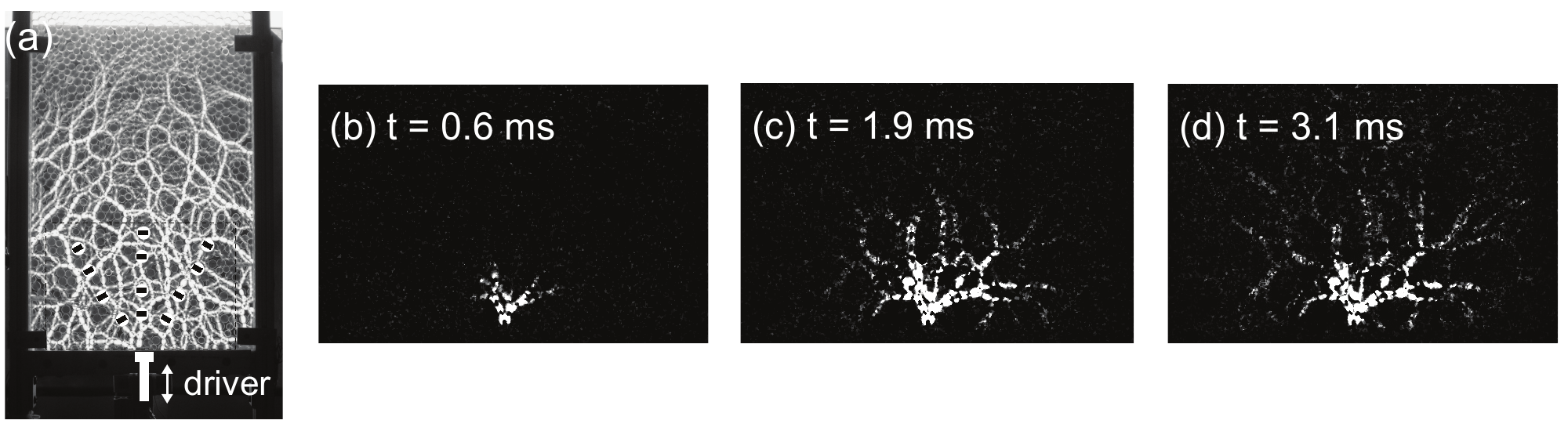}
\caption{(a) Image of photoelastic aggregate viewed through a pair of circular polarizers. (b-d) Representative $|\Delta I|$ measured from region shown in (a), with white representing the areas with largest change in intensity. A movie of the dynamics is available at http://nile.physics.ncsu.edu/pub/movies/gransound/}
\label{fig:setup}
\end{figure*}

Hertzian contact theory underlines models of sound propagation, whether as the interaction potential between particles in discrete element simulations \cite{Cundall-1979-DNM} or in the calculation of an effective bulk modulus in EMT \cite{Digby-1981-EEM}. The contact force $f$ between two particles is given by an equation of the form $f \propto \delta^\beta$, where $\delta$ is the distance each particle is compressed and the exponent $\beta$ depends on the particle geometry. Two common idealized situations are $\beta=3/2$ for spheres and $\beta=1$ for cylinders. The contact area $a$ between the particles, which in EMT are permanently bonded, has a force dependence given by $a \propto f^{\frac{1}{\beta}}$, for a circular contact area in the case of spheres, and $a \propto f^{\frac{1}{2 \beta}}$ for a rectangular contact area in the case of cylinders.  The sound velocity $c$ is governed by the stiffness $s = df/d\delta$ of the contacts; for Hertzian contacts, this is $s \propto \delta^{\beta-1} \propto f^{\frac{\beta-1}{\beta}}$. In experiments on one-dimensional chains of identical spheres, \cite{Coste-1999-VHC} observed quantitative agreement with the expected $c \propto \sqrt{s}$ for a variety of materials.

However, due to the force chain network present within granular systems, the contact force, area, and stiffness are in general heterogeneous quantities, and measured acoustic signals arise from a superposition of signals traveling within this heterogeneous medium. Furthermore, contacts between particles may open and close during the propagation \cite{Schreck-JPS-2010}, in contrast with the well-bonded assumption of EMT. Nonetheless, measurements of sound speed $c$ as a function of system pressure $P$ have observed the expected $c \propto P^{1/6}$ for 2D and 3D systems of spheres above a critical $P$ for which non-Hertzian effects such particle asperities and contact rearrangements are minimized \cite{Goddard-1990-NEP, Jia-1999-UPE, Gilles-2003-LFB}. However, nonlinear and non-affine effects prevent EMT from providing a complete treatment of the dynamics \cite{Makse-2004-GPN}.

An additional complication lies in the observation that $c$ can take different values within the same system, depending on whether the speed measurement is taken for time-of-flight, peak response, or harmonic excitation \cite{Liu-1993-SGM}. From this property of $c$, as well as the observation that a minute expansion of a single particle can cause large shifts in the acoustic response,  \cite{Liu-1993-SGM, Liu-1994-SPS} proposed that the underlying force chain network is at the heart of the failures of EMT. Simulations by \cite{Somfai-2005-EWP} directly addressed the heterogeneity and local dynamics using discrete element simulations of soft spheres. In a 2D system of idealized Hertzian disks ($\beta = 1$), they observed that the propagating wave was in fact coherent and thereby insensitive to the force chain network. They interpret this result as arising from the amplitude and speed both being independent of the local contact forces, as might be the case if the stiffness of the contacts were constant throughout the system for the special case of $\beta=1$. Reconciling the failures of EMT with this unexpected insensitivity to the force chain network requires measurements of the propagation dynamics in real systems at length scales comparable to the particles and force chains. 

In order to address the particular role played by the inter-particle contacts and forces, we make spatiotemporally-resolved measurements in an experimental system in which changes in the force chain network are rendered visible. Our apparatus is filled with a two-dimensional (2D) aggregate of photoelastic \cite{Shukla-1991-DPS, Howell-1999-SF2} disks, allowing direct visualization of the sound path via polarizing filters and a high-speed camera. In addition, piezoelectric sensors embedded in a subset of the particles provide the higher temporal resolution necessary for speed measurements. This combination of techniques allows us to directly measure force, sound amplitude, and sound speed on the particle scale. We observe that the force chain network plays a key role in sound propagation, with the largest amplitude sound propagating along the strongest force chains. We find that a simple model of increasing contact area with inter-particle force provides semi-quantitative agreement with the observed trend. In addition, we find the sound speed of the leading edge of the signal in our 2D system is consistent with the speed found along a 1D chain. However, the 2D speed of the peak response is not observed to be force dependent, and is slower than the corresponding 1D speed. We use this result to explain the speed difference observed for 2D and 3D systems \cite{Liu-1993-SGM} by postulating that the leading edge of the signal travels only along the stiffest (fastest) path, while the peak response is a combination of sound that traveled along multiple paths and would therefore be independent of the force and slower than the leading edge of the signal.

\section{Experiment} 

Our apparatus consists of a $(29 \times 45)$~cm$^2$ vertical aggregate of ${\cal O}(10^3)$ photoelastic disks confined in a single layer between two sheets of Plexiglass (see fig.~\ref{fig:setup}a). A pair of left and right circularly polarizing filters on opposites sides form a polariscope, so that particles with larger force appear brighter and with more fringes. The particles are cut from Vishay PhotoStress material PSM-4, with diameters $d_1 = 9$~mm and $d_2 = 11$~mm in equal concentrations ($\bar{d} \equiv 10$~mm); the zero-frequency Young's modulus of the material is $E_0 =4$~MPa and the density is $\rho=1.06$~g/cm$^3$. Because the particles are viscoelastic, they have a stress-dependent dynamic modulus $E(\omega)$; for the 750 Hz signals used here, $E \approx 50$ to $100$~MPa \cite{modulus}. The upper surface of the aggregate is free and we perform measurements near the bottom, where the Janssen effect \cite{Janssen-1895-VGS, Sperl-2006-ECP} eliminates vertical gradients in the average pressure. The average pressure in this region is $10^{-3}E_0$, with strong force chains typically $10^{-2}E_0$, and the system has an average packing fraction of $0.84 \pm 0.01$  To improve statistics, we perform experiments in 19 different particle configurations.

A voice coil affixed to the bottom wall of the system sinusoidally drives a flat platform of width $2 \bar{d}$, providing a point-like source of acoustic waves. The driver is mechanically isolated from the apparatus to eliminate transmission of signals through the boundaries. The source pulse consists of $5$ consecutive sine waves with a frequency of $750$~Hz, which corresponds to a wavelength $\lambda = 10 \bar d$ to $20 \bar d$ (based on sound speed measurement presented below). Prior to each experiment, we send an annealing sequence of $30$ pulses so that the system settles into a state for which we observe repeatable measurements.  

\begin{figure}
\centering
\includegraphics[width=1\columnwidth]{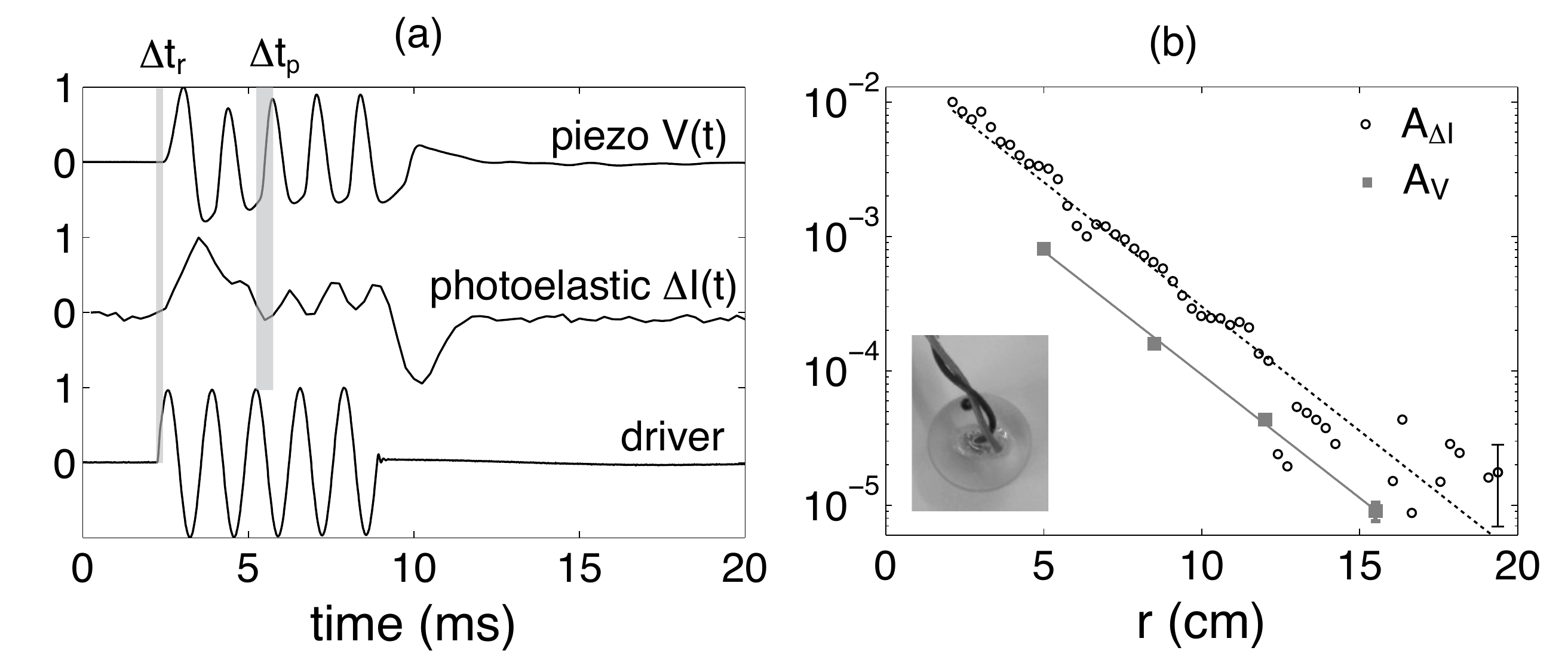}
\caption{Sample signals showing correspondence between two measurement techniques. (a) Driver current (bottom), $\Delta I(t)$ for a representative photoelastic particle (middle) and piezoelectric voltage $V(t)$ (top), with all values normalized to same scale.  Gray bars illustrate the intervals $\Delta t_r$ and $\Delta t_p$. (b) Decay in average signal amplitude with distance from driver:  $A_{\Delta I}(r)$ for the photoelastic signal ($\bigcirc$) and $A_V(r)$  for the piezoelectric signal ($\blacksquare$). Solid lines show $A \propto \exp^{-r/\xi}$ with $\xi = 2.3$~cm.  Inset: particle with embedded piezoelectric sensor}
\label{fig:calibration}
\end{figure}

We observe the wave propagation through the aggregate using two techniques: the photoelastic response via a digital high speed camera, and the electrical response of particle-scale piezoelectric sensors embedded in a subset of particles (see fig ~\ref{fig:calibration}b). The former provides data at all particle locations, but only 4kHz temporal resolution; the latter provides data at only 12 particles but 100 kHz temporal resolution. In addition, a 3k~$\times$~2k digital camera provides higher spatial resolution images for measurement of particle positions and forces.

We image the photoelastic response of the aggregate using a Phantom V5.2 camera operating at $4$~kHz with $512 \times 512$  resolution. Since the change in the photoelastic signal during the pulse is weak, maximally $\pm 10$ out of 256 levels, we transmit the 5-wave pulse 50 times at $20$~ms intervals to allow the previous pulse to dissipate. The frame-by-frame average intensity $I(x,y; t)$ of these 50 pulses is used to measure spatially-resolved dynamics. In each frame, we determine the locations where compression is present by examining differences in the image brightness with respect to the initial frame, $\Delta I(x,y; t) \equiv I(x,y; t_i) - I(x,y; t_{0})$, as shown in fig.~\ref{fig:setup}b-d. The field $A_{\Delta I}(x,y; t)$ records the maximum amplitude observed at each particle location.  Note that for a measured speed of $c=200$~m/s, the 4~kHz frame rate averages the photoelastic response over a distance of approximately $5 \bar d$ in the direction of propagation.

In addition, we utilize embedded piezoelectric sensors to provide higher temporal resolution for a subset of the particles. Piezoelectrics were chosen because they can fit within a single particle, produce a voltage proportional to the applied stress, and have a frequency response well above the frequencies we measure. Our sensors are 2~mm thick lead zirconate titanate piezoelectric ceramics from Piezo Systems, Inc., cut to 4 mm $\times$ 4 mm squares; this adds $30\%$ to each particle's mass.  The configuration of the 12 sensor-particles is shown in fig.~\ref{fig:setup}a, with the first row 5~cm from the driver and the remaining rows spaced 3.5~cm apart. Since our piezoelectric sensors are only sensitive to stresses applied across their thickness, it is important to orient them so that their face is perpendicular to a line connecting the center of the driver with the center of the sensor. The voltage measured on any particles not in this orientation is corrected with the appropriate cosine factor during post-processing. 

During each train of pulses, we record the time-varying voltage produced by each of the twelve piezoelectric sensors (see fig.~\ref{fig:calibration}a). Due to the high impedance of the piezoelectric sensors, an op-amp buffer and a $10$~M$\Omega$ resistor are connected in parallel with the piezoelectric to lower the impedance of the circuit and reduce signal crosstalk. Three quantities are measured for each sensor and the driver: amplitude $A_V$, rise time $t_r$, and the middle peak time $t_p$. We measure amplitude $A_V$ by fitting a sine function to the middle three peaks in $V(t)$. We measure $t_r$ as the time at which each signal rises $15\%$ above its resting value, and $t_p$ as the time at which the maximum value of the middle peak is attained for each signal.  The intervals $\Delta t_r$ and $\Delta t_p$ are measured for each sensor with respect to the driver signal, and are used to calculate the speed of sound $c$ along the observed sound path. 

The two amplitude measurements, photoelastic $A_{\Delta I}(x,y; t)$  and piezoelectric $A_V(x,y; t)$ provide complementary information: the former with better resolution in $(x,y)$ and the latter with better resolution in $t$. As can be seen in fig.~\ref{fig:calibration}a, $\Delta I(t)$ for a single particle shows the expected 5-peaked signal sent by the driver and measured at a nearby sensored particle. In fig.~\ref{fig:calibration}b, for all 19 configurations, we average $A_V(r)$ over the 3 sensors in each row and azimuthally average $A_{\Delta I}(r)$. We find that $\langle A_V (r)\rangle$ and $\langle A_{\Delta I}(r) \rangle$ exhibit the same exponential decay. As such, the two amplitudes provide the same information. Note that this exponential decay is due to the dissipative nature of both the viscoelastic photoelastic polymer and the frictional granular contacts, and suppresses the multiply-scattered (coda) response \cite{Liu-1994-SPS, Jia-2004-CMS}.

These particle-scale measurements of sound amplitude and propagation speed can be correlated with the measured particle-scale forces in order to quantify the effect of the force network on the propagation dynamics. In order to find the force $F$ on each particle, we locate all particles on a high-resolution image and calculate the squared average intensity gradient $|\nabla I|^2$  within each particle. This empirical method \cite{Howell-1999-SF2} is calibrated to known forces in a separate test cell. Three calibration regimes were used for low-, mid-, and  high-force particles.

\section{Results} 

\subsection{Contact Force Law}
 
We directly measure the contact force law for our particles by compressing a single particle between two plates with the force applied along the particle's diameter. fig.~\ref{fig:linearity}a shows the measured force for a range of applied displacements. Within a range which corresponds to the typical forces in the full experimental system, we find that the contact force law is approximately of the form
\begin{equation}
  f \propto \delta^{5/4}.
  \label{forcelaw}
\end{equation}
While this deviates from the Hertzian prediction that co-planar cylinders follow a linear contact law, we note that slight out-of-plane rotations of the particles lead to imperfect alignment at their contacts. The value $\beta = 5/4$ can therefore be interpreted as lying between $\beta=1$ (for rectangular contacts between co-planar cylinders) and $\beta=3/2$ (for circular contacts between perpendicular cylinders) \cite{Johnson}. As will be seen below, this deviation from a linear contact law has several detectable consequences.

\subsection{Linearity and nonlinearity}

\begin{figure}
\centering
\includegraphics[width=.9\columnwidth]{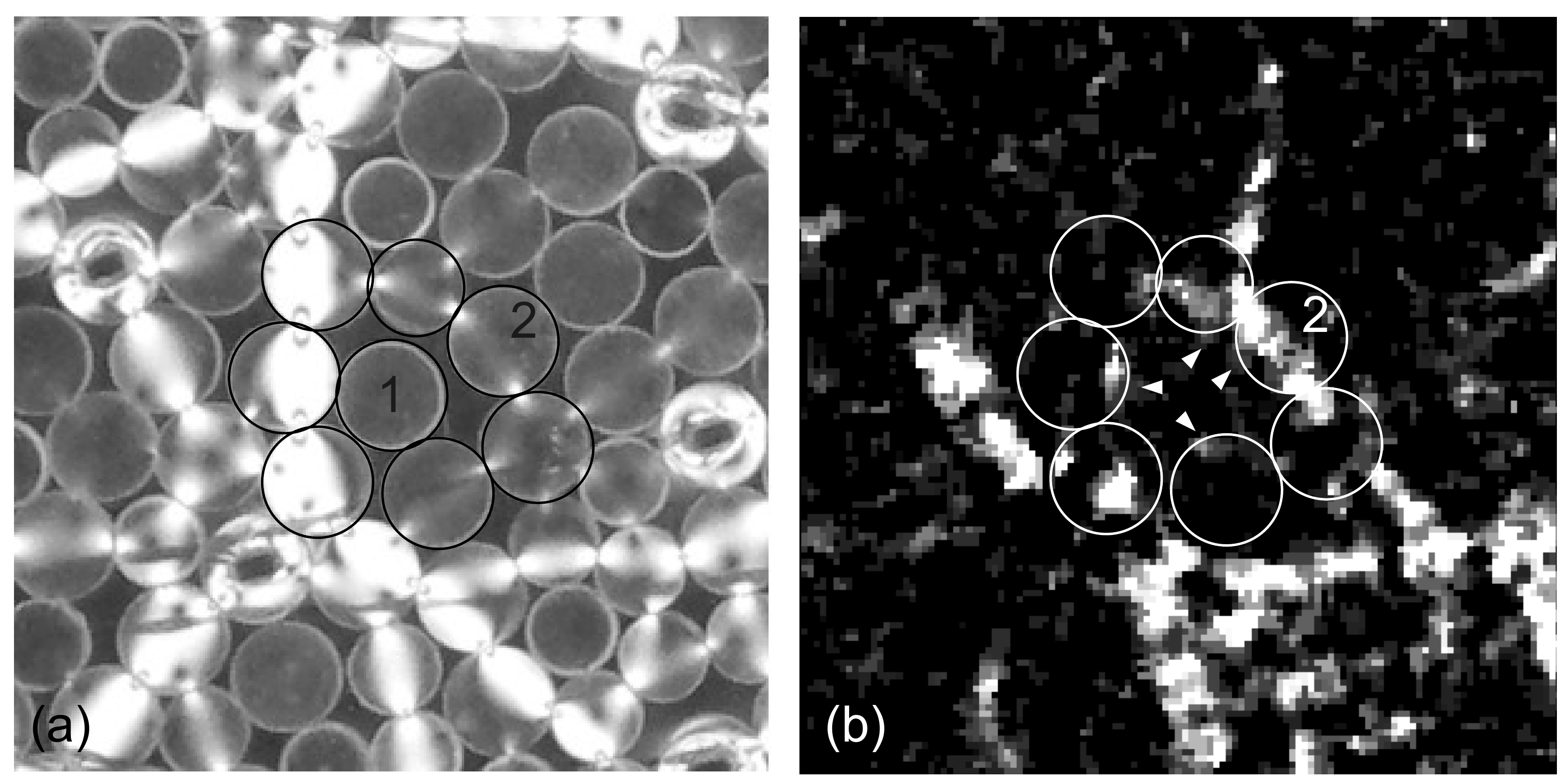}
\caption{(a) Detail of force chains approximately $8 \bar d$ from the driver, with particles 1 and 2 exhibiting weak contacts. (b) $|\Delta I|$ during sound propagation for the same region. New contacts to particle 1 ($\blacktriangleright$) and the new chain through particle 2 open and close repeatedly during each wave cycle.}
\label{fig:transient}
\end{figure}

The photoelastic particles allow us to directly visualize an important particle-scale nonlinearity present in the system. Due to the non-zero Poisson ratio of the polymer material ($\nu \approx 0.5$), a particle under dynamic compression expands in the perpendicular direction, potentially creating new (but transient) contacts during the transmission of a wave. The photoelastic particles render the creation of such new mechanical contacts visible, as can be seen in fig.~\ref{fig:transient}. In the static image of the force chains (a), particles 1 and 2 are not part of the strong force chain network. However, during the transmission of the pulse (b) each of these marked particles exhibits a measurable $|\Delta I|$ before returning to its initial, un-stressed state. This observation that particle contacts can change during sound propagation violates the well-bonded assumption of effective medium theory \cite{Digby-1981-EEM} because the connectivity of the network transiently changes due to the passage of the wave. Similar, but permanent, changes to the force network were presumably at work in experiments where minute thermal expansion \cite{Liu-1993-SGM} or consolidation of loose packs \cite{Hostler-2005-PWP} modified the signal coda.

\begin{figure}
\centering
\includegraphics[width=1\columnwidth]{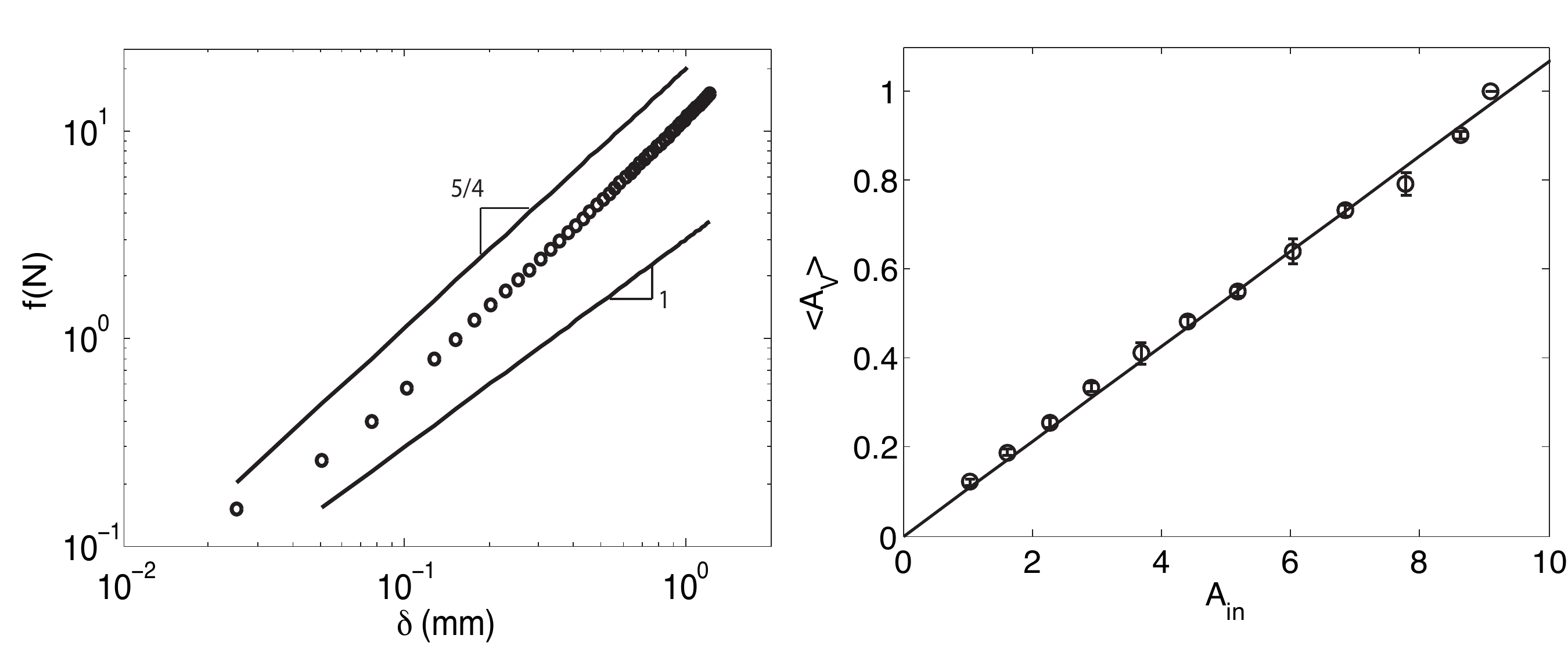}
\caption{(a) Measured contact force law compared with $f \propto \delta^{5/4}$ and  $f \propto \delta$ (solid lines). (b) $\langle A_V \rangle$ found from all twelve piezoelectric sensors in a single pack as a function $A_{in}$.  $A_V$ and $A_{in}$ are in arbitrary units of pressure.  Error bars are standard error and the solid line shows $A_V \propto A_{in}$.}
\label{fig:linearity}
\end{figure}

In spite of the inherent nonlinearity of such events, we nonetheless observe an approximately linear response in the average propagation amplitude of the system. As shown in fig.~\ref{fig:linearity}b, we measure the input amplitude $A_{in}$ as the pressure generated by the driver as it compresses the particles, and the output amplitude as the average signal  $\langle A_{V} \rangle$ received by all 12 piezoelectric sensors in a particular configuration. Since the piston is driven sinusoidally, it has a maximum displacement of $\delta=\frac{F_{coil}}{\omega^2 m}$, where $F_{coil}$ is the electrical force from the coil (measured via the current), $\omega$ is the driving frequency, and $m$ is the mass of the piston plus some effective mass of the system. For a given $\omega$, system configuration, and $F_{coil}$, this unknown mass remains constant. Therefore, $\delta \propto F_{coil}$ is not only the displacement of the piston, but also the compression applied to the first particle. Using this relation to provide $\delta$ in eq.~\ref{forcelaw}, we find that the applied pressure $A_{in} \propto F_{coil}^{5/4}$. As we increase the current to the coil, we can thereby measure  $\langle A_{V} \rangle$ as a function of $A_{in}$ and find an approximately linear relationship. This suggests that at the driving magnitudes investigated here, the transient contacts are not a large enough effect to detectably destroy an approximately linear response.

\subsection{Amplitude}

\begin{figure}
\centering
\includegraphics[width=0.7\columnwidth]{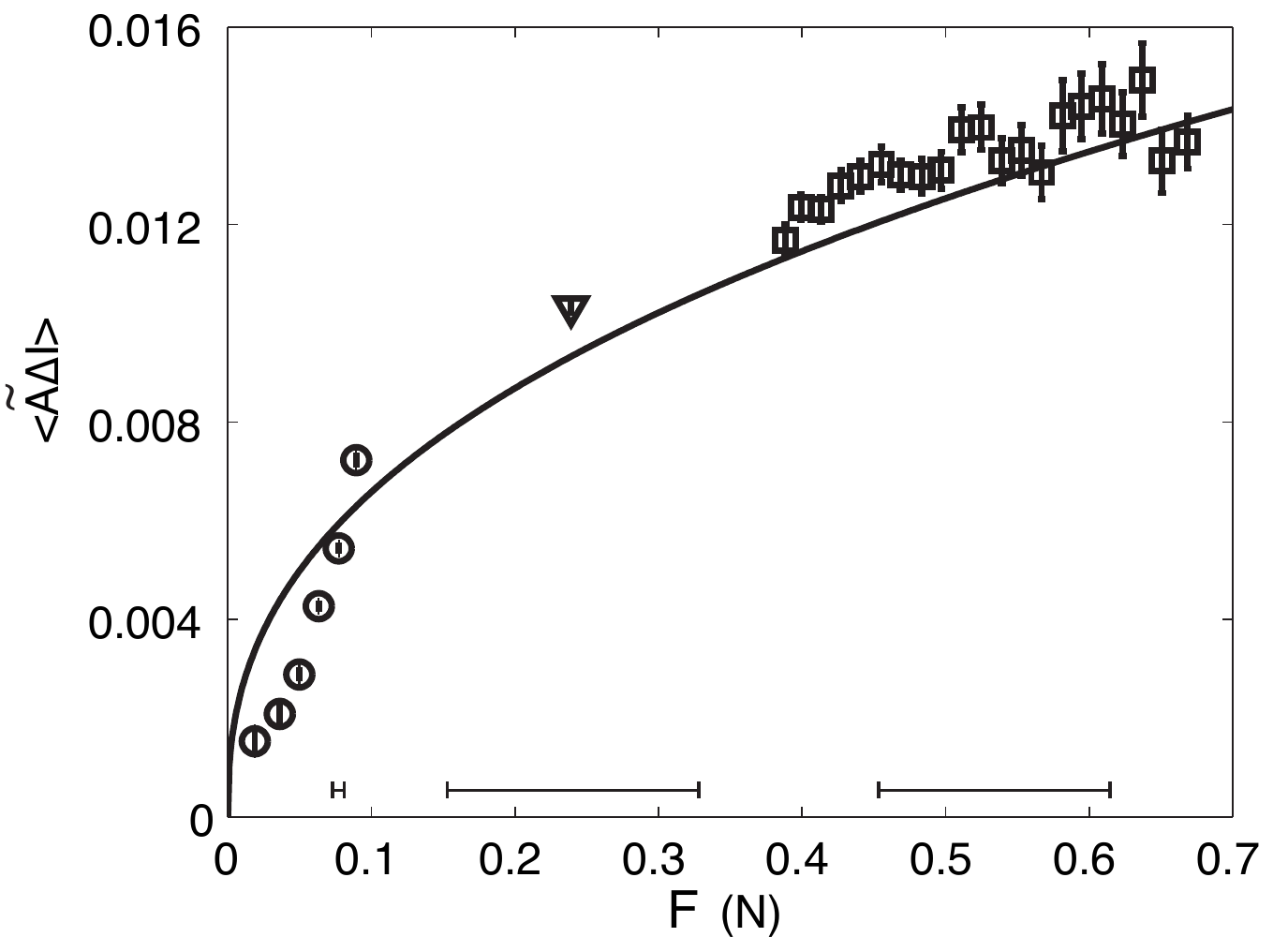}
\caption{ Average scaled sound amplitude $\langle \tilde A_{\Delta I} \rangle$ as a function of particle force $\bar F$, with each data point averaged over several hundred measurements. Symbols ($\bigcirc$,  $\triangledown$,  $\square$) are for calibrations in the low-, mid-, and high-force regimes, respectively, with error bars for individual measurements in each regime shown at bottom of plot.  Line is $A \propto F^{2/5}$.}
\label{fig:amplitude}
\end{figure}

As can be seen in fig.~\ref{fig:setup}b-d, the dynamic response is quite similar in structure to the original force chain structure shown in fig.~\ref{fig:setup}a. To quantify this effect, we measure both the local force $F$ on each particle and the maximum amplitude $A_{\Delta I}$ of the wave passing through that particle. For all particles in each of the 19 configurations, we bin the particles by $F$ and calculate the average $\langle \tilde A_{\Delta I} \rangle$, where $\tilde A$ is corrected for the exponential decay shown in fig.~\ref{fig:calibration}b. As can be seen in fig.~\ref{fig:amplitude}, the particles with stronger forces on average experience a larger amplitude, which corresponds to the visual observation.

We can understand this amplitude trend in terms of the measured contact law in eq.~\ref{forcelaw}. The Hertzian contact area $a$ formed between two such disks pushed together by a pair of opposing forces $f$ is $a \propto f^{2/5}$.  If the contacts between particles are the main conduits for sound propagation, then we would also expect the magnitude of the sound to vary as $\tilde A_{\Delta I} \propto F^{2/5}$. This relationship is consistent with the data shown in fig.~\ref{fig:amplitude}, particularly outside the low-force regime. This contact model is an over-simplification in several important ways. While the Hertzian contact theory was developed for diametric loading, the multiple contacts on each particles arise at various angles. Furthermore, while the sound propagates radially outward from the driver, the force chains form arcs which are not in general aligned with this direction (see fig.~\ref{fig:setup}). As such, a more sophisticated treatment would  consider only the radial component of the stress tensor. The fact that we recover an approximate scaling of the amplitude points to the robustness of the trend.

\subsection{Sound Speed}

\begin{figure}
\centering
\includegraphics[width=0.7\columnwidth]{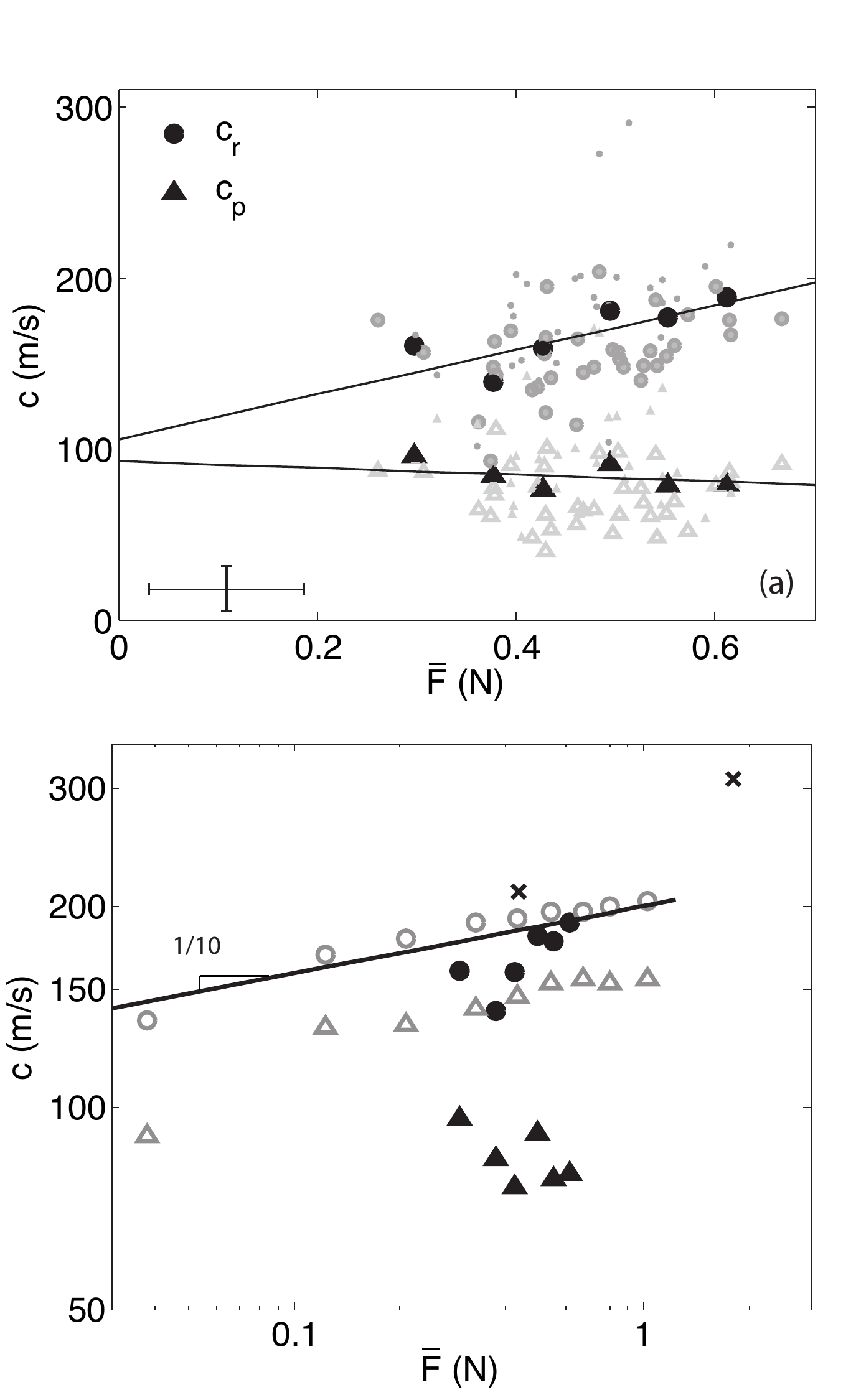}
\caption{(a) Sound speed measured by the piezoelectric sensors as a function of the average force $\bar F$ along the observed path. Large symbols are for sensors at $r=5$~cm, small symbols at $r=8.5$~cm. The speed of the initial rise $c_r$ (gray $\bullet$) and of the peak $c_p$ (gray $\blacktriangle$) with linear fits to each. Black symbols are averages of the same data, binned by force. Linear fits are: 
$c_r(\bar F) = (106 \pm 46)$~m/s $ + (130 \pm 97)$~s/kg $\bar F$
$c_p(\bar F)  =  (93 \pm 38)$~m/s $ - ( 20 \pm 80)$~s/kg $\bar F$  
with a confidence interval of $95\%$. 
(b) Comparison of the 1D (gray) and 2D (black, copied from (a)) speeds, $c_r$ ($\bullet$) and $c_p$ ($\bigtriangleup$), with $c \propto f^{1/10}$ for comparison. Bulk speeds ($\times$) are calculated from measured dynamic modulus under fixed compression \cite{modulus}.} 
\label{fig:speed}
\end{figure}

We measure sound speed via the signal travel time from the driver to the piezoelectric sensors (see fig.~\ref{fig:calibration}a) along the dominant propagation path from the driver to each sensor from the high-speed movies (see fig.~\ref{fig:setup}b-d). As in \cite{Liu-1993-SGM, Somfai-2005-EWP}, we record both the travel time for the initial rise ($\Delta t_r$) and for the peak response ($\Delta t_p$). The distance travelled is taken to be the {\itshape observed path} (rather than Cartesian distance) since this is the path over which we observed the largest amplitude sound propagates. This path is selected by hand from the movies; we measure the path length $L$ and calculate the two sound speeds, $c_r = L/\Delta t_r$ and $c_p = L/\Delta t_p$. In addition, we measure the average force $\bar F$ from the average force within the particles along the observed path. Due to the exponential decay in the amplitude of the signals, quantitative speed data is only available for the inner two rows of sensors located at $r \lesssim \lambda$.

In fig.~\ref{fig:speed}a, we observe that $c_r$ is on average faster than $c_p$ (as was previously observed in \cite{Liu-1993-SGM, Somfai-2005-EWP}, see Discussion for the significance of this result), $c_r$ and $c_p$ are both independent of distance from the driver as expected, and that there is significant scatter in the data which impedes our interpretation of the force-dependence. As most of the scatter in fig.~\ref{fig:speed}a is due to uncertainty in the photoelastic force measurement, we bin the data by force in order to clarify the force-dependence, and compare it with both  Hertzian predictions and  speed measurements in one-dimensional (1D) chains. 

We performed sound speed measurements on 1D chains of the $11$~mm particles confined by a known force within a linear channel centered above the driver. For Hertzian contacts, we expect $c \propto \sqrt{s} \propto f^{\frac{\beta-1}{2 \beta}} \propto f^{1/10}$, which provides reasonable agreement with the measured sound speeds, $c_r$ and $c_p$, on the 1D chain shown in fig.~\ref{fig:speed}b. Notably, these speeds are faster than the bulk speed predicted by the zero-frequency Young's modulus ($c = \sqrt{E_0/\rho} = 61$~m/s). However, $E$ depends on both the driving frequency and degree of pre-loaded stress due to viscoelastic stiffening. Independent measurements of the dynamic $E$ at typical values of $\omega$ and $f$ \cite{modulus}, predict a speed consistent with what we observe.

The agreement of the 1D and 2D sound measurements of $c_r$ (see fig.~\ref{fig:speed}b) suggests that the Hertzian contact model provides a good description of the weak force-dependence. This observation contrasts with numerical simulations using ideal ($\beta=1$) Hertzian disks and Coulombic friction, where \cite{Somfai-2005-EWP} observed no such force-dependence in either speed or amplitude measurements. We attribute this difference to our measured force law (eq.~\ref{forcelaw}) which has $\beta = 5/4$; in a 3D aggregate of Hertzian spheres ($\beta=3/2$), similar force-dependence might be expected. Additionally, we note that in 2D $c_p$ is slower than $c_p$ in 1D and is approximately independent of the force (see Discussion below.)

\section{Discussion} 

These experiments are performed on a highly dissipative viscoelastic material within a near-field regime ($r \ll \lambda$). Nonetheless, we are able to reproduce features seen in more conventional 3D granular materials ($c_r > c_p$) and 1D chains ($c \propto \sqrt{s}$) which indicate a degree of universality among systems of quite different dimensionality, particle geometry, and material properties. In part, this is due to the fact that the viscoelasticity only affects the constant of proportionality in $f \propto \delta^\beta$,  while $\beta$ depends only on the contact geometry. However, it is possible that such additional stiffening enhances the importance of the force chains. In spite of these limitations, the use of these photoelastic materials permits us to address the importance of particle-scale heterogeneities to sound propagation. The application of these techniques outside the near-field regime will require the use of less-dissipative materials.

As shown in fig.~\ref{fig:amplitude}, we observe that the local sound amplitude increases with the force on the particle, with a shape that suggests the the effect is due to increasing contact area ($A \propto a \propto f^{2/5}$). This finding is in contrast with a lack of amplitude-dependence in simulations of idealized ($\beta=1$) disks. \cite{Somfai-2005-EWP}. For realistic deformable particles (rather than merely overlap with a repulsive potential), the contact area should still increase with force: $a \propto f^{\frac{1}{2 \beta}} \propto f^{1/2}$.

In experiments on sand \cite{Liu-1993-SGM} and in numerical simulations \cite{Somfai-2005-EWP}, the measured value of $c$ depended on the method: the speed at the leading edge of the signal, $c_r$, was observed to be several times faster than that of the peak response, $c_p$. By definition, $c_r$ represents the arrival of the first signal, that which arrived along the fastest (stiffest) path. Through the use of photoelastic particles to measure the dominant sound path, we observe that these fast/stiff paths correspond to force chains. Because $c_r$ is sensitive to the particular force chain configuration through $s \propto f^\frac{\beta - 1}{\beta}$, it shows evidence of (weak) force dependence (see fig.~\ref{fig:speed}b). In contrast, the peak response measured by $c_p$ is composed of signals which have travelled via many different paths of varying stiffness and amplitude. Because this portion of the wave has traveled on many different paths, the speed is on average independent of the force along the fastest path, as seen in fig.~\ref{fig:speed}a. This may be related to simulations which show $c_p$ saturating to a maximum value as a function of $r$ \cite{Mouraille-2006-SWA}. Also note that $c_p$ is faster in 1D chains than in the 2D system, since all signals must travel on the same particle-scale path. The remaining speed deficit is thereby due to either dispersion or path-effects at the sub-particle scale.

Thus, prior observations of $c_r > c_p$ are likely due to $c_r$ measuring sound that travelled along the (stiffer) force chain network. A consequence of the force-chain sensitivity of $c_r(f)$ is that the leading edge of the wave front will not form a coherent arc propagating outward, but will instead be irregular. The behavior along each segment of force chain may be reminiscent of sound propagation in 1D granular chains \cite{Coste-1999-VHC}, but with a network of connections between the chains. Interestingly, in the limit of zero confining pressure (no force chains), the observation that the intercepts of $c_r(f)$ and $c_p(f)$ are approximately equal would appear to suggest that the compressional wave speed $c$ takes a single value in this limit. However, in studies of granular acoustics near this jamming transition, additional nonlinear effects are present which are relevant to understanding the scaling of shear and compressional waves \cite{Bonneau-2008-ERH, Jacob-2008-APJ}.

\section{Conclusions} 

In conclusion, we observe that both the amplitude and speed of sound are influenced by the force chains, and that these chains may change due to the transmission of the sound pulse. This highlights the role of force chains as more than just a contact network between grains.  Local properties of the material are crucial to local sound propagation dynamics, and mean field models therefore are unlikely to capture the full behavior of the system.  The dependence of these results on the wavelength of the disturbance \cite{Kondic-2009-PDG}, either above or below the length scale set by the force chains, may provide a means to acoustically probe the force chain network in non-photoelastic 3D systems.

\section{Acknowledgements}

 The authors would like to thank Chiara Daraio, Paul Johnson, Ellak Somfai, and Nathalie Vriend for helpful discussions, and Cliff Chafin and Stephanie Couvreur for development work on the project. This work was supported by the National Science Foundation (DMR-0644743). 


\begin{thebibliography}{10}

\expandafter\ifx\csname url\endcsname\relax\def\url#1{\texttt{#1}}\fi

\bibitem{Digby-1981-EEM}
\Name{Digby P.~J.} \REVIEW{Journal of Applied Mechanics}{48}{1981}{803}.


\bibitem{Goddard-1990-NEP}
\Name{Goddard J.~D.} \REVIEW{Proceedings Of The Royal Society Of London Series
  A}{430}{1990}{105}.

\bibitem{Velicky-2002-PDS}
\Name{Velicky B. \and Caroli C.} \REVIEW{Physical Review E}{65}{2002}{021307}.


\bibitem{Jia-1999-UPE}
\Name{Jia X., Caroli C. \and Velicky B.} \REVIEW{Physical Review
  Letters}{82}{1999}{1863}.


\bibitem{Makse-1999-WEM}
\Name{Makse H.~A., Gland N., Johnson D.~L. \and Schwartz L.~M.}
  \REVIEW{Physical Review Letters}{83}{1999}{5070}.


\bibitem{Makse-2004-GPN}
\Name{Makse H., Gland N., Johnson D. \and Schwartz L.} \REVIEW{Physical Review
  E}{70}{2004}{061302}.

\bibitem{Zeravcic-2008-LBV}
\Name{Zeravcic Z., van Saarloos W. \and Nelson D.~R.} \REVIEW{EPL (Europhysics
  Letters)}{83}{2008}{44001}.

\bibitem{Xu-2009-ETJ}
\Name{Xu N., Vitelli V., Wyart M., Liu A. \and Nagel S.} \REVIEW{Physical
  Review Letters}{102}{2009}{038001}.


\bibitem{Wyart-2010-SPT}
\Name{Wyart M.} \REVIEW{EPL (Europhysics Letters)}{89}{2010}{}.

\bibitem{Cundall-1979-DNM}
\Name{Cundall P.~A. \and Strack O. D.~L.} \REVIEW{Geotechnique}{29}{1979}{47}.

\bibitem{Coste-1999-VHC}
\Name{Coste C. \and Gilles B.} \REVIEW{European Physical Journal
  B}{7}{1999}{155}.


\bibitem{Schreck-JPS-2010}
\Name{Schreck C.~F., Bertrand T., O'Hern C.~S. \and Shattuck M.~D.} \REVIEW{arXiv:1012.0369v1}{}{2010}.

\bibitem{Gilles-2003-LFB}
\Name{Gilles B. \and Coste C.} \REVIEW{Physical Review Letters}{90}{2003}.

\bibitem{Liu-1994-SPS}
\Name{Liu C.~H.} \REVIEW{Physical Review B}{50}{1994}{782}.

\bibitem{Liu-1994-SVG}
\Name{Liu C.~H. \and Nagel S.~R.} \REVIEW{Journal of Physics -- Condensed
  Matter}{}{1994}{A433}.

\bibitem{Somfai-2005-EWP}
\Name{Somfai E., Roux J.~N., Snoeijer J.~H., {van Hecke} M. \and {van Saarloos}
  W.} \REVIEW{Physical Review E}{72}{2005}{021301}.

\bibitem{Shukla-1991-DPS}
\Name{Shukla A.} \REVIEW{Optics and Lasers in Engineering}{14}{1991}{165}.

\bibitem{Howell-1999-SF2}
\Name{Howell D., Behringer R.~P. \and Veje C.} \REVIEW{Physical Review
  Letters}{82}{1999}{5241}.

\bibitem{modulus}
Measurements made by TechSource Engineering, Inc., Erie, PA.

\bibitem{Janssen-1895-VGS}
\Name{Janssen H.~A.} \REVIEW{Zeitschr. d. Vereines deutscher   Ingenieure}{39}{1895}.

\bibitem{Sperl-2006-ECP}
\Name{Sperl M.} \REVIEW{Granular Matter}{8}{2006}{59}.

\bibitem{Jia-2004-CMS}
\Name{Jia X.} \REVIEW{Physical Review Letters}{93}{2004}{15}.

\bibitem{Johnson}
\Name{Johnson K.~L.} \Book{Contact Mechanics} (Cambridge University Press)
  1987.

\bibitem{Liu-1993-SGM}
\Name{Liu C.~H. \and Nagel S.~R.} \REVIEW{Physical Review B}{48}{1993}{15646}.

\bibitem{Hostler-2005-PWP}
\Name{Hostler S.~R. \and Brennen C.~E.} \REVIEW{Physical Review
  E}{72}{2005}{031303}.


\bibitem{Mouraille-2006-SWA}
\Name{Mouraille O., Mulder W.~A. \and Luding S.} \REVIEW{Journal of Statistal
  Mechanics}{}{2006}{P07023}.

\bibitem{Bonneau-2008-ERH}
\Name{Bonneau L., Andreotti B. \and Cl\'{e}ment E.} \REVIEW{Physical Review
  Letters}{101}{2008}{118001}.

\bibitem{Jacob-2008-APJ}
\Name{Jacob X., Aleshin V., Tournat V., Leclaire P., Lauriks W. \and Gusev V.}
  \REVIEW{Physical Review Letters}{100}{2008}{158003}.

\bibitem{Kondic-2009-PDG}
\Name{Kondic L., Dybenko O.~M. \and Behringer R.~P.} \REVIEW{Physical Review
  E}{}{2009}{}.

\end{thebibliography}

\end{document}